\documentstyle[preprint,aps,floats,tighten]{revtex}

\newcommand{\etal}{{\it et al.}}
\newcommand{\mt}{m_T}
\newcommand{\ppbar}{p\overline{p}}
\newcommand{\Dzero}{D\O}
\newcommand{\dzero}{\Dzero~Collaboration}
\newcommand{\mw}{M_{W}}
\newcommand{\mz}{M_{Z}}
\newcommand{\qqbar}{q\overline{q}}
\newcommand{\pt}{p_T}
\newcommand{\Et}{E_T}
\newcommand{\met}{\mbox{${\hbox{$E$\kern-0.6em\lower-.1ex\hbox{/}}}_T \!$ }}
\newcommand{\mpt}{\mbox{$\rlap{\kern0.1em/}\pt$}}
\newcommand{\pe}{p(e)}
\newcommand{\pev}{\vec\pe}
\newcommand{\pte}{\pt(e)}
\newcommand{\ptev}{\vec\pte}
\newcommand{\utv}{\vec\ut}
\newcommand{\ut}{u_T}
\newcommand{\ptnu}{\pt(\nu)}
\newcommand{\ptnuv}{\vec\ptnu}
\newcommand{\ptw}{\pt(W)}
\newcommand{\ptwv}{\vec\ptw}
\newcommand{\wev}{W\to e\nu}
\newcommand{\zee}{Z\to ee}
\newcommand{\wte}{W\to \tau\nu\to e\nu\overline\nu\nu}
\newcommand{\cem}{c_{\rm EC}}
\newcommand{\lt}{<}
\newcommand{\gt}{>}
\newcommand{\alphaem}{\alpha_{\rm EC}}
\newcommand{\deltaem}{\delta_{\rm EC}}

\newcommand{\srec}{s_{\rm rec}}
\newcommand{\ptee}{\pt(ee)}
\newcommand{\pteev}{\vec\ptee}

\newcommand{\GEAN}{{\sc geant}}

\newcommand{\PM}{\pm}

\newcommand{\phie}{\phi(e)}
\newcommand{\PRL}{Phys. Rev. Lett.}
\newcommand{\PL}{Phys. Lett.}
\newcommand{\PR}{Phys. Rev.}

\newcommand{\NIM}{Nucl. Instrum. Methods in Phys. Res.}
\newcommand{\ZP}{Z.~Phys.}

\newcommand{\cdf}{CDF~Collaboration}

\def\lsim{\mathrel{\rlap{\lt \lower4pt\hbox{\hskip1pt$\sim$}}}}
\begin{document}
\pagestyle{myheadings}
\addtolength{\textheight}{26pt} 
\title{ A Measurement of the $W$ Boson Mass Using Electrons at Large
 Rapidities}
%
\author{                                                                      
B.~Abbott,$^{46}$                                                             
M.~Abolins,$^{43}$                                                            
V.~Abramov,$^{19}$                                                            
B.S.~Acharya,$^{12}$                                                          
I.~Adam,$^{45}$                                                               
D.L.~Adams,$^{53}$                                                            
M.~Adams,$^{29}$                                                              
S.~Ahn,$^{28}$                                                                
V.~Akimov,$^{17}$                                                             
G.A.~Alves,$^{2}$
% ...
\\                                                                            
\vskip 0.30cm                                                                 
\centerline{(D\O\ Collaboration)}                                             
\vskip 0.30cm                                                                 
}                                                                             
\address{                                                                     
\centerline{$^{1}$Universidad de Buenos Aires, Buenos Aires, Argentina}       
\centerline{$^{2}$LAFEX, Centro Brasileiro de Pesquisas F{\'\i}sicas,         
                  Rio de Janeiro, Brazil}                                     
\centerline{$^{3}$Universidade do Estado do Rio de Janeiro,                   
                  Rio de Janeiro, Brazil}                                     
\centerline{$^{4}$Institute of High Energy Physics, Beijing,                  
                  People's Republic of China}                                 
\centerline{$^{5}$Universidad de los Andes, Bogot\'{a}, Colombia}             
\centerline{$^{6}$Universidad San Francisco de Quito, Quito, Ecuador}         
\centerline{$^{7}$Institut des Sciences Nucl\'eaires, IN2P3-CNRS,             
                  Universite de Grenoble 1, Grenoble, France}                 
\centerline{$^{8}$Centre de Physique des Particules de Marseille,             
                  IN2P3-CNRS, Marseille, France}                              
\centerline{$^{9}$DAPNIA/Service de Physique des Particules, CEA, Saclay,     
                  France}                                                     
\centerline{$^{10}$Panjab University, Chandigarh, India}                      
\centerline{$^{11}$Delhi University, Delhi, India}                            
\centerline{$^{12}$Tata Institute of Fundamental Research, Mumbai, India}     
\centerline{$^{13}$Kyungsung University, Pusan, Korea}                        
\centerline{$^{14}$Seoul National University, Seoul, Korea}                   
\centerline{$^{15}$CINVESTAV, Mexico City, Mexico}                            
\centerline{$^{16}$Institute of Nuclear Physics, Krak\'ow, Poland}            
\centerline{$^{17}$Institute for Theoretical and Experimental Physics,        
                   Moscow, Russia}                                            
\centerline{$^{18}$Moscow State University, Moscow, Russia}                   
\centerline{$^{19}$Institute for High Energy Physics, Protvino, Russia}       
\centerline{$^{20}$Lancaster University, Lancaster, United Kingdom}           
\centerline{$^{21}$University of Arizona, Tucson, Arizona 85721}              
\centerline{$^{22}$Lawrence Berkeley National Laboratory and University of    
                   California, Berkeley, California 94720}                    
\centerline{$^{23}$University of California, Davis, California 95616}         
\centerline{$^{24}$University of California, Irvine, California 92697}        
\centerline{$^{25}$University of California, Riverside, California 92521}     
\centerline{$^{26}$Florida State University, Tallahassee, Florida 32306}      
\centerline{$^{27}$University of Hawaii, Honolulu, Hawaii 96822}              
\centerline{$^{28}$Fermi National Accelerator Laboratory, Batavia,            
                   Illinois 60510}                                            
\centerline{$^{29}$University of Illinois at Chicago, Chicago,                
                   Illinois 60607}                                            
\centerline{$^{30}$Northern Illinois University, DeKalb, Illinois 60115}      
\centerline{$^{31}$Northwestern University, Evanston, Illinois 60208}         
\centerline{$^{32}$Indiana University, Bloomington, Indiana 47405}            
\centerline{$^{33}$University of Notre Dame, Notre Dame, Indiana 46556}       
\centerline{$^{34}$Purdue University, West Lafayette, Indiana 47907}          
\centerline{$^{35}$Iowa State University, Ames, Iowa 50011}                   
\centerline{$^{36}$University of Kansas, Lawrence, Kansas 66045}              
\centerline{$^{37}$Kansas State University, Manhattan, Kansas 66506}          
\centerline{$^{38}$Louisiana Tech University, Ruston, Louisiana 71272}        
\centerline{$^{39}$University of Maryland, College Park, Maryland 20742}      
\centerline{$^{40}$Boston University, Boston, Massachusetts 02215}            
\centerline{$^{41}$Northeastern University, Boston, Massachusetts 02115}      
\centerline{$^{42}$University of Michigan, Ann Arbor, Michigan 48109}         
\centerline{$^{43}$Michigan State University, East Lansing, Michigan 48824}   
\centerline{$^{44}$University of Nebraska, Lincoln, Nebraska 68588}           
\centerline{$^{45}$Columbia University, New York, New York 10027}             
\centerline{$^{46}$New York University, New York, New York 10003}             
\centerline{$^{47}$University of Rochester, Rochester, New York 14627}        
\centerline{$^{48}$State University of New York, Stony Brook,                 
                   New York 11794}                                            
\centerline{$^{49}$Brookhaven National Laboratory, Upton, New York 11973}     
\centerline{$^{50}$Langston University, Langston, Oklahoma 73050}             
\centerline{$^{51}$University of Oklahoma, Norman, Oklahoma 73019}            
\centerline{$^{52}$Brown University, Providence, Rhode Island 02912}          
\centerline{$^{53}$University of Texas, Arlington, Texas 76019}               
\centerline{$^{54}$Texas A\&M University, College Station, Texas 77843}       
\centerline{$^{55}$Rice University, Houston, Texas 77005}                     
\centerline{$^{56}$University of Washington, Seattle, Washington 98195}       
}                                                                             
\maketitle
         \centerline{ \today}             
   
\vskip -0.1cm
\begin{abstract}
We report a measurement of the $W$ boson mass based on an integrated
luminosity of 82 pb$^{-1}$ from $\ppbar$ collisions at $\sqrt{s}=1.8$~TeV
recorded  in 1994--1995 by the \Dzero\ detector at the Fermilab Tevatron. We
identify $W$ bosons by their decays to $e\nu$, where the electron is 
 detected in the forward calorimeters. We extract the mass by fitting
the transverse mass and  the electron and neutrino transverse momentum 
spectra of 11{,}089 $W$ boson candidates. 
 We measure $\mw=80.691\pm0.227$~GeV. By combining this
 measurement with our previously published central  
calorimeter results from data 
 taken in 1992--1993 and 1994--1995, we obtain $\mw=80.482\pm0.091$~GeV. 
\end{abstract}
\addtolength{\textheight}{-26pt}
\narrowtext

In the standard model of the electroweak interactions (SM), the mass
of the $W$ boson is predicted to be 
\begin{equation}
\mw = \left( \frac{\pi\alpha(\mz^2)}
 {\sqrt{2}G_F}\right)^\frac{1}{2} \frac{1}{\sin\theta_W\sqrt{1-\Delta r}}\; .
\label{eq:dr}
\end{equation}
In the ``on-shell'' scheme~\cite{on_shell} $\cos\theta_W = \mw/\mz$, 
where $\mz$ is the $Z$ boson mass.  A measurement of $\mw$,
together with $\mz$~\cite{wworld}, 
 the Fermi constant ($G_F$), and the electromagnetic 
coupling constant ($\alpha$),
 experimentally determines the weak radiative corrections $\Delta r$.
Compared to the formulation in~\cite{on_shell} where $\alpha$ was defined at
$Q^2=0$, we have absorbed  purely electromagnetic corrections 
 into the value of $\alpha$
by evaluating it at $Q^2=\mz^2$. 
The dominant SM contributions to $\Delta r$
arise from loop diagrams involving the top quark and the Higgs boson. 
If additional particles coupling to the $W$ boson exist, they also give
contributions to $\Delta r$. Therefore, a measurement of
$\mw$ is a stringent experimental test of SM predictions.
Within the SM, measurements of $\mw$ and the mass of the top quark constrain  
the mass of the Higgs boson.

We report a new measurement of the $W$ boson mass using electrons 
detected at forward angles. We use  82 pb$^{-1}$ of 
 data  recorded with the \Dzero\ detector during the
1994--1995 run of the Fermilab Tevatron $\ppbar$ collider. This forward 
electron measurement, in addition to increasing the
  statistical
 precision, complements our previous
 measurement with central electrons \cite{wmass1bcc} because the more complete
 combined
 rapidity coverage gives useful constraints on model parameters that 
 reduce the systematic error. 
 A more complete account of this measurement can be 
 found in Ref.~\cite{wmass1bec}. 

At the Tevatron, $W$ bosons are produced through $\qqbar ^ \prime$
annihilation. $\wev$ decays  are
characterized by an electron with large transverse energy
($\Et$) and significant transverse momentum imbalance ($\mpt$) 
due to the undetected neutrino. 
The particles 
recoiling against the $W$ boson
are referred to collectively as the ``underlying event.''

The \Dzero\ detector~\cite{d0_nim} consists of three major subsystems: a
  tracking
detector, a calorimeter,  and a muon spectrometer.
The tracking detector consists of  a vertex drift chamber, a central 
 drift chamber (CDC), and two forward drift
chambers (FDC). The CDC covers the pseudorapidity ($\eta = - \ln \left( \tan
    \frac{\theta}{2} \right)$ where $\theta$ is the polar angle)
 region $|\eta|<1.0$.
The FDC extend the coverage to $|\eta|<3.0$. The  central calorimeter
(CC) and two end calorimeters (EC) 
 provide almost uniform coverage for particles with $|\eta|<4$.

At the trigger level, we require $\mpt>15$~GeV and an energy cluster in the
electromagnetic (EM) calorimeter with $\Et>20$~GeV.
The cluster must be isolated and have a shape consistent with that of an 
electron shower.

During event reconstruction, electrons are identified as energy clusters in the
EM calorimeter, which satisfy isolation and shower shape cuts and
have a drift chamber track pointing to the cluster centroid.
We determine forward electron  energies by adding the 
energy depositions in the  calorimeter within a cone
  of radius 20 cm, centered on
the  cluster centroid. 
The electron momentum ($\pev$) is 
determined by combining its energy with the direction 
obtained from the shower 
 centroid position and the drift chamber track.
The trajectory of the electron defines the position of the
event vertex along the beamline. 

We measure the sum of the transverse momenta
of all particles recoiling against the $W$ boson, $\utv = \sum_i E_i
\sin\theta_i \hat u_i^T$, where $E_i$ is the energy deposition in 
 calorimeter cell $i$, $\hat u_i^T$ is the unit transverse vector 
 pointing from the beamline to the cell center, and $\theta_i$ is the polar
 angle defined by the cell 
center and the event vertex. 
The $\utv$ calculation  excludes  cells occupied by the electron.
The transverse momenta of the neutrino, 
$\ptnuv = -\ptev-\utv$, and the $W$ boson,
$\ptwv=-\utv$, are inferred from momentum conservation. 

We select a $W$ boson sample of 11{,}089 events by requiring 
$\ptnu>30$~GeV, $\ut<15$~GeV, and an electron candidate with 
$1.5<|\eta|<2.5$ and $\pte>30$~GeV.

We extract the $W$ boson mass from the spectra of the electron $\pte$, 
 neutrino  $\ptnu$, 
and the transverse mass, $\mt = \sqrt{2\pte\ptnu(1-\cos\Delta\phi)}$,
where $\Delta\phi$ is the azimuthal separation between the two leptons.
For each spectrum 
 we perform a maximum likelihood fit to the data using probability density
functions from a Monte Carlo program.
We model
the production dynamics of $W$ bosons and the detector response to
  predict the spectra.
The $\mt$, $\pte$, and $\ptnu$ spectra
have quite different sensitivities to 
 the $W$ boson production dynamics and the recoil
momentum measurement. By performing the measurement using all three spectra we
 provide a powerful cross-check with complementary systematics.

$Z$ bosons decaying to electrons provide an important control
sample. We use them to calibrate the
detector response to the underlying event and electrons, and to 
constrain the model for vector boson production used in the
Monte Carlo simulations.  
We trigger on $Z\rightarrow ee$ events having at least 
 two EM clusters with $\Et\gt 20$~GeV. 
We accept $Z\rightarrow ee$ decays with at least one forward electron with
$1.5<|\eta|<2.5$ and  another forward or central
 ($|\eta|<1.0$) electron. A  
 central electron is required 
to have $\pt>25$~GeV but is allowed not to have a matching drift chamber 
track. The forward electron candidate is required to have  $\pt>30$~GeV and 
a matching drift chamber track. 
 This selection accepts 1{,}687 $Z$ boson events.

 We use a fast Monte Carlo program developed  for the 
  central electron analyses  \cite{wmass1bcc,wmass1acc}, with  
 modifications in the simulation of forward electron events. The program 
generates $W$ and $Z$ bosons with the
$\eta$ and $\pt$ spectra given by a calculation~\cite{ly} which 
 used soft gluon 
resummation and the MRST~\cite{mrst} parton distribution 
functions. 
We use the relativistic Breit-Wigner line shape with mass-dependent width,
 skewed by the mass
dependence of the parton luminosity. The measured $W$ and $Z$ boson intrinsic 
widths~\cite{width} are used.
The angular distribution of the decay electrons includes a $\ptw$-dependent 
${\cal O}(\alpha{_s}^{2})$ correction~\cite{mirkes}. The program also generates
$\wev\gamma$~\cite{berends}, $\zee\gamma$~\cite{berends}, and $\wte$ decays.

The program smears the generated $\pev$ and $\utv$ vectors using a
parameterized  detector response model and applies inefficiencies introduced by
the trigger  and offline selection requirements. Backgrounds are added to the 
 Monte Carlo samples. The parameters are adjusted to match the data.

The electron energy resolution  ($\delta{E}/E$) is parameterized by 
 calorimeter sampling,
noise, and constant terms. In the Monte Carlo simulation of forward electrons 
we use a sampling term
of $15.7\%/\sqrt{E/\hbox{GeV}}$, derived from beam tests~\cite{d0ec}.
 The noise term is
determined by pedestal distributions taken from the $W$ boson
 data. We
constrain the constant term to $\cem = 1.0^{+0.6}_{-1.0}\%$ by requiring that
the predicted width of the dielectron invariant mass
  spectrum be consistent with the $Z$ boson data.

\begin{figure}[htb]
\begin{tabular}{c}
\includegraphics{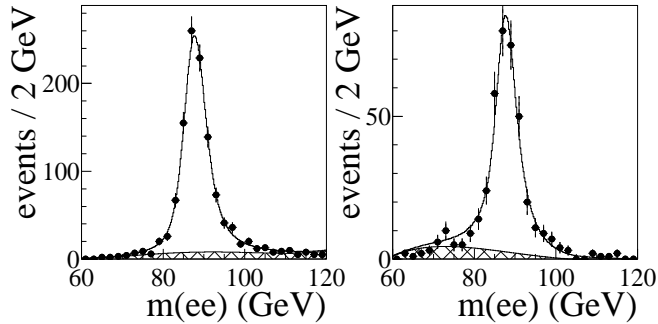}
\end{tabular}

\caption[]{The dielectron invariant mass distribution of the CC/EC
 (left, $\chi^2$/dof = $14/19$) and EC/EC
  (right, $\chi^2$/dof = $12/17$) $Z$ boson data  ($\bullet$). 
  The solid line shows the fitted signal plus background
shape and the small hatched area shows the background. The fitting window is
  $70 < m(ee) < 110$ GeV. }
\label{fig:mee}
\end{figure}

Beam tests show that the electron energy response of the end
 calorimeter can be
parameterized by a scale factor $\alphaem$ and an offset $\deltaem$. We
determine these {\it in situ} using 
$\zee$ decays~\cite{wmass1bec}. 
For forward electrons we obtain $\deltaem = -0.1\pm0.7$~GeV and 
$\alphaem =0.95179\pm0.00187$ by fitting the observed mass spectra
while constraining the resonance masses to the $Z$ boson mass. 
The uncertainty on $\alpha_{EC}$ is dominated by the finite size of the $Z$
boson 
 sample. Figure~\ref{fig:mee} shows the observed mass spectra from
the dielectron samples and the line shapes 
 predicted by the Monte Carlo simulation for the 
fitted values of $\cem$, $\alphaem$, and $\deltaem$. The background was 
 determined from a sample of events with two EM clusters failing 
 electron quality cuts. 
The calibration 
of the electron polar angle~\cite{wmass1bec} uses muons
  from $\ppbar$ collisions and cosmic rays to calibrate the
drift chambers, and $\zee$ decays to align the EC  
 with the drift chambers. 

\vskip 1.5 in

We calibrate the response of the detector to the underlying event using the
$Z$ boson data sample. 
In $\zee$ events, momentum conservation requires  
$\pteev=-\utv$, where $\pteev$ is the sum of the two 
 electron $\pt$ vectors~\cite{wmass1bec}.
We constrain the detector response $R_{\rm rec}$ using the mean 
value of the $\pteev+\utv$ projection on the inner bisector
  of the two electron directions.
 $Z$ boson  
 events with two  
forward electrons give a recoil response measurement  that's consistent
  with the measurement performed in 
 the central dielectron analysis~\cite{wmass1bcc}. 

\begin{figure}[htb]
\begin{tabular}{ll}
\includegraphics{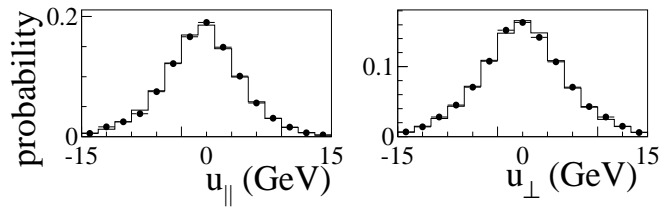}
\end{tabular}
\caption[]{ Probability distributions of $u_{\parallel}$ 
 (left, $\chi^2$/dof = $25/15$) and $u_{\bot}$ (right, $\chi^2$/dof =
  $14/15$) for the forward 
 $W$ boson data ($\bullet$) and the Monte Carlo simulation~(---).}
\label{fig:hadres}
\end{figure}

The recoil momentum resolution has two components: a stochastic term,
 which we model as $\srec/\sqrt{\pt/\hbox{GeV}}$; and
  the detector noise and pile-up, which we model using the 
 scaled $\mpt$ from  random $\ppbar$ interactions~\cite{wmass1bec}.
 We constrain the model by comparing the observed rms of 
$\pteev+\utv/R_{\rm rec}$ with Monte Carlo predictions.
The model tuned for the central electron analysis~\cite{wmass1bcc} gives
 a good description of the $\pteev+\utv/R_{\rm rec}$  
 distributions for our $Z$ 
 boson  event sample. 
Figure~\ref{fig:hadres} shows the comparison between the $W$ boson Monte Carlo
  and the  data of the projection of recoil momentum on the
direction of the forward electron ($u_{\parallel}$) and on a direction
   perpendicular to the electron momentum ($u_{\bot}$). 

\vskip 1.0in

\begin{figure}[htb]
\begin{tabular}{cc}
\includegraphics{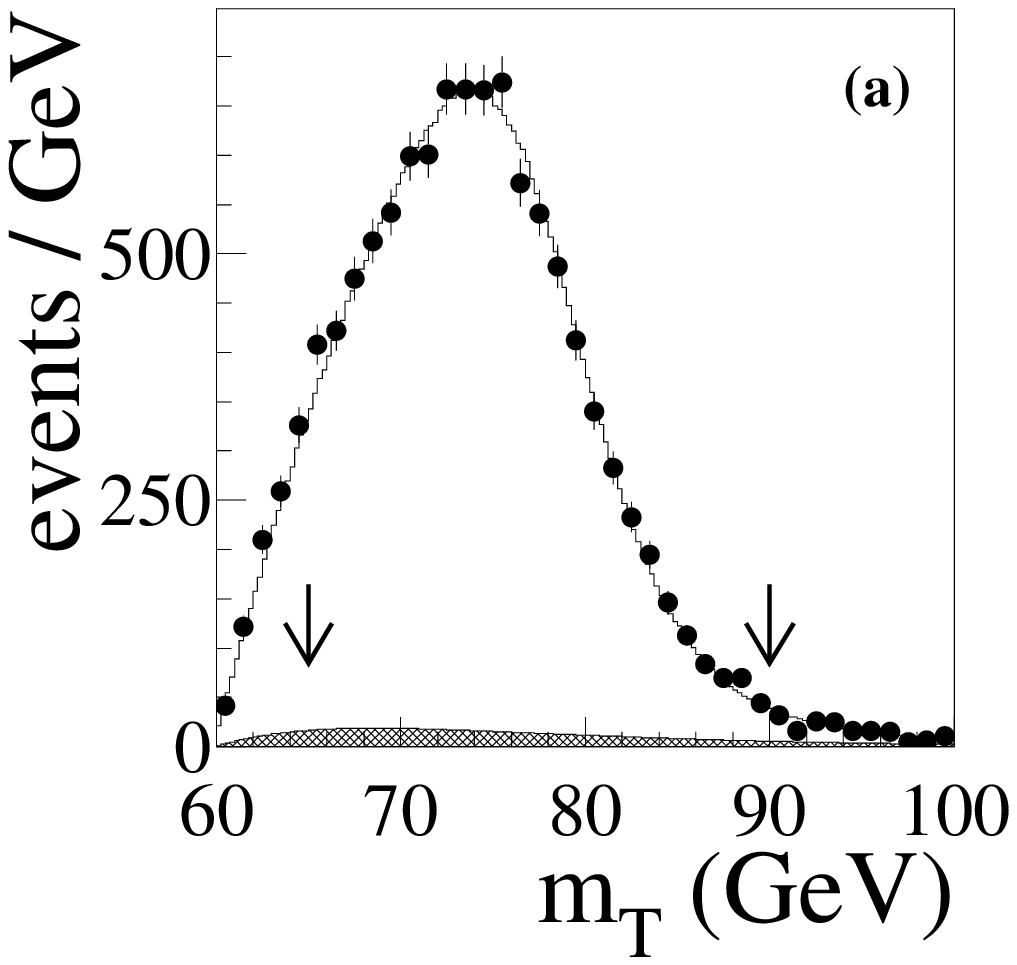}
\includegraphics{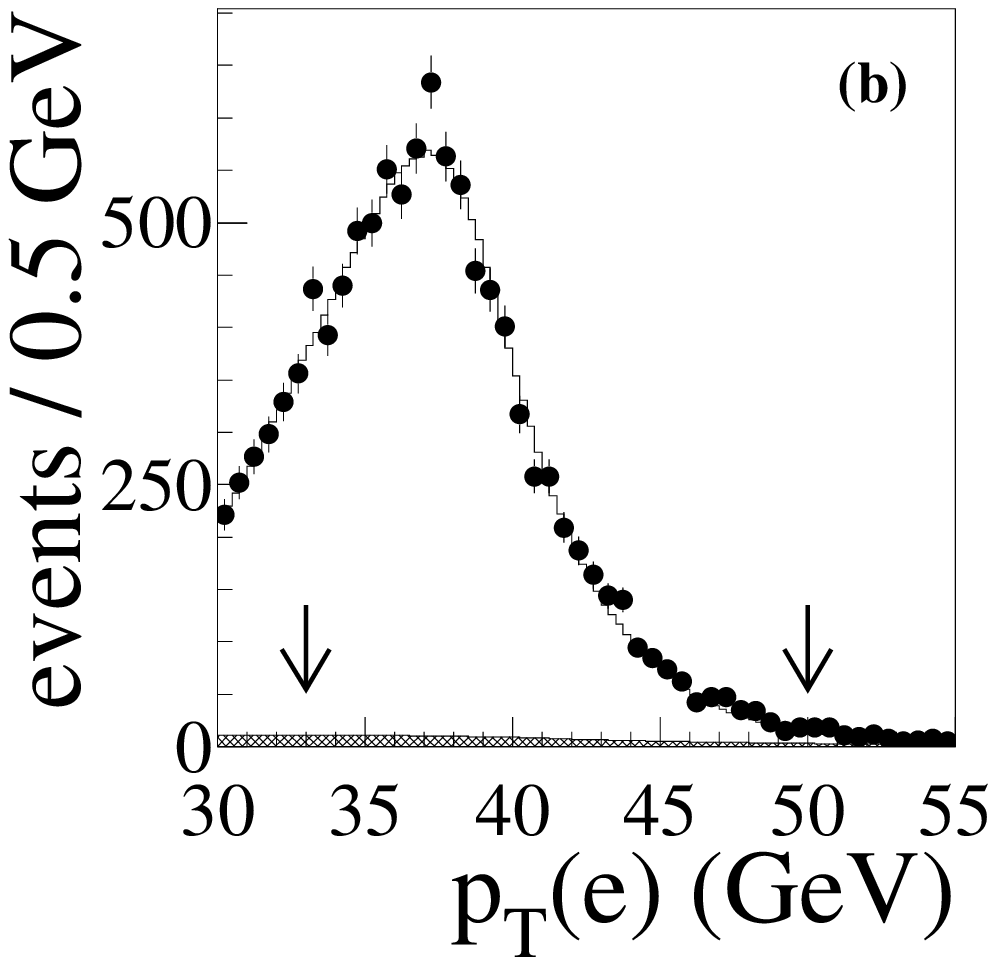}
\includegraphics{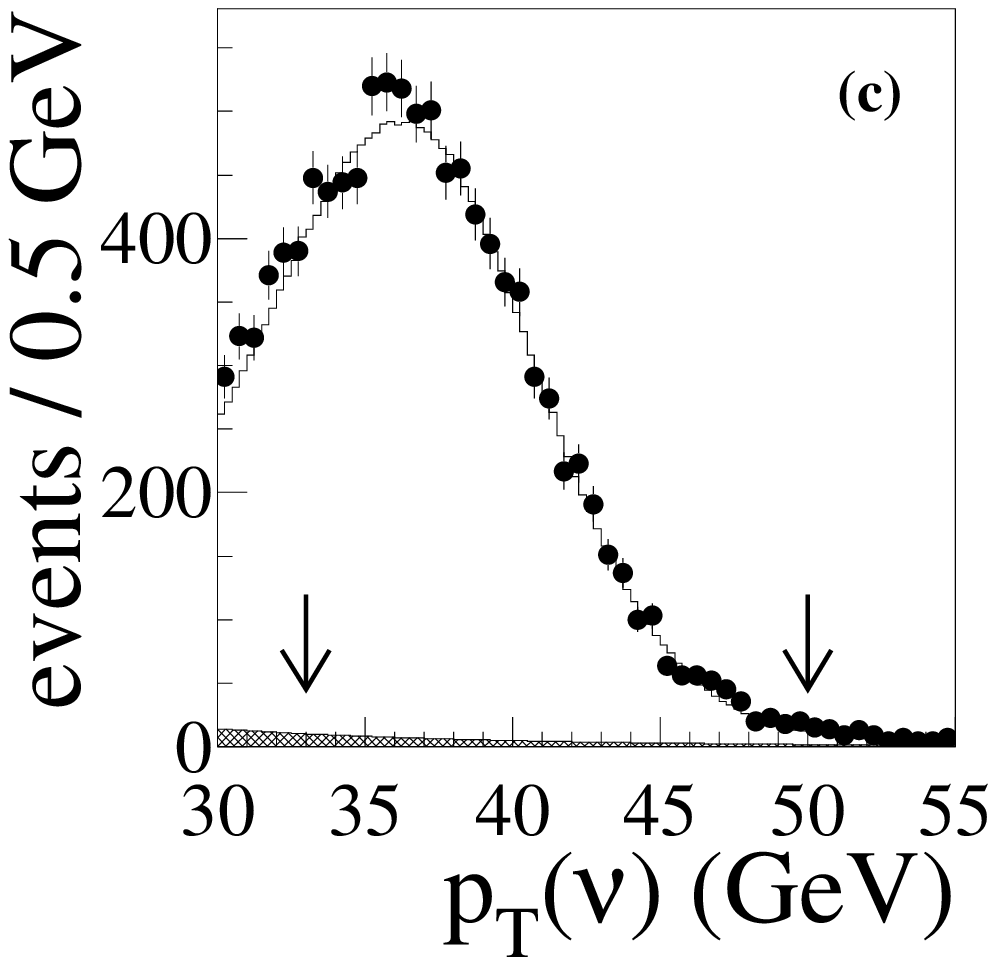}
\end{tabular}
\vskip 0.5cm
\caption[]{Spectra of \\ (a) $\mt$, (b) $\pte$, and \\ (c) $\ptnu$
 from the \\ data ($\bullet$), 
the fit (---), and \\ the backgrounds (shaded).\\ The arrows
indicate the fit \\windows.}
\label{fig:mt_fit}
\end{figure}

Backgrounds in the $W$ boson sample are due to $\wte$ decays 
 (1\%, included in the Monte Carlo simulation), 
hadrons misidentified as electrons (3.64$\PM$0.78~\%, determined from the 
 data) and $\zee$ decays 
(0.26$\PM$0.02~\%, determined from {\sc herwig}~\cite{herwig} and
 \GEAN\ \cite{geant}
 simulations).
  Their shapes
  are included in the probability density functions used in the fits.
 The results of the fits to the $\mt$, $\pte$, and $\ptnu$  distributions 
 are shown in Fig.~\ref{fig:mt_fit} and  Table~I. 

 We estimate the systematic uncertainties in $\mw$ (Table~II) 
 by varying  the Monte Carlo parameters
 within their uncertainties.
We assign an uncertainty that
characterizes the range of variations in $\mw$ obtained when employing
several recent parton distribution functions: 
 MRST, MRS(A$^\prime$)~\cite{mrsa}, 
 MRSR2~\cite{mrsr2}, CTEQ3M~\cite{cteq3m}, CTEQ4M~\cite{cteq4m} and 
 CTEQ5M~\cite{cteq5m}. We have 
 checked that the pdf's reproduce the $\eta(e)$ distribution
   for the $W$ bosons well~\cite{wmass1bec}.
 We allow 
 the $\ptw$ spectrum to vary within constraints derived from the $\ptee$ 
 spectrum
of the $Z$ boson
 data~\cite{wmass1bcc} and from $\Lambda_{\rm QCD}$~\cite{wmass1bcc}. 
Smaller uncertainties in $\mw$ are due to
the removal of the cells occupied by the electron from the computation of 
$\utv$, and the 
modeling of trigger and selection biases~\cite{wmass1bec}. The 
uncertainty due to radiative decays contains an estimate of the effect of
neglecting double photon emission in the Monte Carlo 
simulation~\cite{baur_twophoton}.

The total systematic errors are shown in Table~I.
The good agreement of the three fits shows that our simulation models 
the $W$ boson production dynamics and the detector response well.
Fits to the data in bins of
luminosity, $\phie$, $\eta(e)$, and $\ut$ and with changes to the fit window 
 show no evidence of  systematic biases.  

\vskip 1.5in
\vskip 0.2in

\begin{table}[ht]
\begin{center}
\caption{The fitted values and errors of the forward 
 $W$ boson mass measurements in GeV. 
 The confidence level (C.L.) is given by the $\chi^2$ probability of the fit. }
\vskip 0.1cm
\begin{tabular}{lccccr}
\hline
Fit                   & Mass & Stat. &Syst.& Total Error & C.L. \\
\hline
$\mt$ fit             & 80.757 & 0.107 & 0.204 &  0.230 &  81\% \\
\hline
$\pte$ fit            & 80.547 & 0.128 & 0.203 &  0.240 & 8\%\\
\hline
$\ptnu$ fit           & 80.740 & 0.159 & 0.310 &  0.348 & 33\% \\
\hline
\end{tabular}
\end{center}
\end{table}

\begin{figure}[htb]
\begin{tabular}{l}
\includegraphics{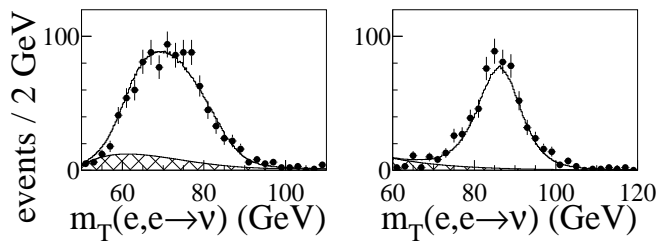}
\end{tabular}
\caption[]{Spectra of the $Z$ boson
 transverse mass from the CC/EC data (left) and
the EC/EC data (right). 
 The superimposed curves show the maximum likelihood fits and the hatched 
  regions show the estimated backgrounds. } 
\label{fig:mtz}
\end{figure}

As a consistency check, we fit the transverse mass distribution of the 
$Z \to ee$ events. We retain one electron in the EC and ignore the energy
 of the other electron (in the CC or EC). The fitted $Z$ boson mass 
 (Fig.~\ref{fig:mtz})  is  $92.004\pm0.895$~(stat) GeV for  the
 CC/EC sample, and $91.074\pm0.299$~(stat) GeV for the EC/EC sample.
The combined mass is $91.167\pm0.284$~(stat) GeV.
 These results  are consistent 
with the input $Z$ boson mass we used to calibrate the detector response. 

\begin{table}[ht]
\begin{center}
\caption{Uncertainties in the combined $\mt$, $\pte$, and $\ptnu$ $W$ boson 
 mass measurement in MeV, for the  forward sample (first column),
 and the combined central and forward  1994--1995 sample (second
 column). }
\vskip 0.1cm
\begin{tabular}{lcc}
\hline
Source                      & Forward & Forward + Central\\
\hline
$W$ boson statistics             & 108 & 61  \\
\hline
$Z$ boson statistics             & 181 & 59  \\
\hline
Calorimeter linearity       &  52 & 25  \\
\hline
Calorimeter uniformity      & --  &  8  \\
\hline
Electron resolution         & 42  & 19  \\
\hline
Electron angle calibration  & 20  & 10 \\
\hline
Recoil response             & 17  & 25  \\
\hline
Recoil resolution           & 42  & 25  \\
\hline
Electron removal            & 4   & 12  \\
\hline
Trigger and selection bias      & 5   &  3  \\
\hline
Backgrounds                 & 20  &  9  \\
\hline
Parton distribution functions     & 17  &  7  \\
\hline
Parton luminosity           & 2   &  4  \\
\hline
$\ptw$ spectrum                    & 25  & 15  \\
\hline
$W$ boson width               & 10  & 10  \\
\hline
Radiative decays       & 1   & 12  \\
\hline
\end{tabular}
\end{center}
\end{table}

 We combine  the $\mt$, $\pte$, and $\ptnu$  measurements of $\mw$ 
  using a full
  covariance matrix that takes into account correlations between all the 
 parameters describing the $W$ boson production model and detector
  response, as well as the statistical correlations.  
 The combination of  all three forward electron measurements yields  
  a $W$ boson mass of $\mw = 80.691\pm0.227$~GeV. 
 We also combine  the three central electron measurements \cite{wmass1bcc} with
  the three forward  $W$ boson mass measurements
  to obtain the combined 1994--1995 data  measurement of 
 $\mw = 80.498\pm0.095$~GeV.
  The $\chi^2$ is  5.1/5 dof, with a probability 
 of 41\%. Further combining this with the 
  measurement from the 1992--93 \cite{wmass1acc} data gives the 1992--95 
 data measurement of~$\mw = 80.482\pm0.091$~GeV. 
 This measurement subsumes all previously published measurements of the $W$
 boson mass by \Dzero.

\vskip 1.2in

From Eq.~\ref{eq:dr},
 using $\alpha (\mz^2) =  (128.88\pm0.09)^{-1}$ \cite{PDG} 
 we find $\Delta r  = -0.0322\pm0.0059$, which 
establishes the existence of  loop corrections to $\mw$ at the 
level of five standard deviations. 
Taken together with our  measured top quark mass 
 ($m_t = 172.1 \pm 7.1$~GeV~\cite{d0_top}), our value of the $W$ boson mass
  is consistent with measurements by
 CDF~\cite{cdfwmass} and the LEP experiments~\cite{wworld} and
  with the SM prediction 
 for a low mass Higgs boson (i.e. $m_H \lt 100$~GeV), and is in even better
  agreement with predictions~\cite{susy} in the MSSM framework.

%
We thank the Fermilab and collaborating institution staffs for 
contributions to this work, and acknowledge support from the 
Department of Energy and National Science Foundation (USA),  
Commissariat  \` a L'Energie Atomique (France), 
Ministry for Science and Technology and Ministry for Atomic 
   Energy (Russia),
CAPES and CNPq (Brazil),
Departments of Atomic Energy and Science and Education (India),
Colciencias (Colombia),
CONACyT (Mexico),
Ministry of Education and KOSEF (Korea),
and CONICET and UBACyT (Argentina).

\end{document}